\documentclass[prl,preprint,showpacs,preprintnumbers]{revtex4}
\usepackage{amsmath,amssymb,epsfig}
\begin{document}

\title{Route to ferromagnetism in organic polymers}
\author{Zsolt~Gul\'acsi$^{1,2}$, Arno~Kampf$^{1}$, and
Dieter~Vollhardt$^{1}$}
\address{$^{1}$ Theoretical Physics III, Center for
Electronic Correlations and Magnetism, Institute of Physics,
University of Augsburg, D-86135 Augsburg, Germany \\
$^{2}$ Department of Theoretical Physics, University of
Debrecen, H-4010 Debrecen, Hungary}
\date{\today}
\begin{abstract}
Employing a rigorous theoretical method for the construction of exact many-electron ground states we prove that interactions can be employed to tune a bare dispersive band structure such that it develops a flat band. Thereby we show that pentagon chain polymers with electron densities above half filling may be designed to become ferromagnetic or half metallic.
\end{abstract}
\maketitle


Conducting polymers \cite{Uint1} are a fascinating class of materials with a
strikingly wide range of applications, e.g., in nanoelectronics \cite{Uint2},
nanooptics \cite{Uint3}, and medicine \cite{Uint3m}. Many of them contain
chains of five-membered-rings as a building block. Such pentagon chain
polymers have been explored \cite{Uint4,Uint5} and utilized \cite{Uint2}
intensively in the past.  In particular,  polythiophene
\cite{Uint6,Uint7,Uint8} was studied in the search for plastic ferromagnets
and, more generally, for ferromagnetism in systems made entirely of nonmagnetic
elements. The possibility for ferromagnetism in these systems was investigated
theoretically \cite{Uint9,Uint10}, with a particular focus on ferromagnetism
due to flat electronic bands arising in odd-membered ring structures \cite{MT}.
Particular attention was paid to the role of side groups of the pentagon ring,
since these may cause flat bands in the band structure. Suwa \emph{et al.}  \cite{SuwaPAM} proposed that ferromagnetism in pentagon-chain polymers such as polydimethylaminopyrrole is related to the hybridization of narrow $\sigma$ bands with wide $\pi$ bands and therefore modelled this polymer by a periodic
Anderson model. In the latter model the electronic interaction acts site selective within the unit cell and the choice of this model \cite{SuwaPAM} was an attempt to account for the different atoms on the pentagon chain.

In this Letter we investigate pentagon chain polymers (see Fig. 1) by a general
multi-band Hubbard model where the electrons experience local Coulomb
interactions on all lattice sites. The microscopic parameters are chosen such that they account for the particular environment and type of atom in the unit cell of the material; in particular, in our approach repulsive on-site interactions are permitted to differ on individual sites. Similarly, we also include bond dependent hopping amplitudes. The hopping parameters are not assumed to take special values leading to flat bands in the bare band structure. By contrast we will show rigorously that the dispersion of the correlated system may be \emph{tuned by the interaction} to become flat. Thereby transitions to ferromagnetic states or correlated half-metallic states at high electron densities may be induced. We thus prove by exact means the conjecture of Brocks \emph{et al.} \cite{B15} that the Coulomb interaction is able to stabilize magnetic order in acene and thiophene.

Our analytic approach proceeds in three steps: the transformation of the
Hamiltonian into positive semidefinite form, the construction of ground states,
and the proof of their uniqueness. This technique is independent of the spatial
dimension and does not require integrability of the model. Previously it was
successfully applied to construct exact ground states for Hubbard chains with
other geometrical structures \cite{B23} and even for the three-dimensional
periodic Anderson model \cite{B21}. Details of the method are described in Ref.
\cite{B23}.

\begin{figure}[t!]
\epsfxsize=8.0cm
\centerline{\epsfbox{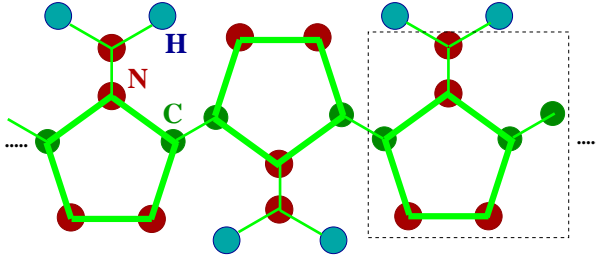}}
\caption{(Color online) Schematic view of the pentagon-chain polymer
polymethylaminotriazole. The dotted square indicates the cell
presented in detail in Fig. 2.}
\label{Fig1}
\end{figure}
Since the analytic technique employed here is applicable to a large class of chains we discuss it in its general form, and then specify it to a model analysis of the organic pentagon chain shown in Fig. 1. The unit cell contains $m=m_p+m_e\geq 2$ sites where $m_p$ is the number of sites in the closed polygon as indicated in Fig. 2. In this case $m_p$=5, and $m_e=1$ is the number
of sites in the side groups; the number $m$ (here $m=6$) also determines the number of sublattices. The i-th unit cell then contains the sites ${\bf i}+{\bf r}_n$, $n \leq m$, with ${\bf r}_1={\bf 0}$ and ${\bf r}_{m+1} ={\bf a}$; here $|{\bf a}|$ is the lattice constant. Altogether the chain consists of $N_c$ cells, where neighboring cells connect through the single point ${\bf r}_{m+1}$. Subsequently we use periodic boundary conditions and fix the number of electrons to $N\leq N_\Lambda$, where $N_\Lambda=mN_c$ is the number of sites. The filling is denoted by $\rho=N/2 N_\Lambda \leq 1$. In the following
$\sum_{\bf i}$, $\prod_{\bf i}$ (or $\sum_{\bf k}$, $\prod_{\bf k}$ in momentum
representation) mean sums and products, respectively, over the $N_c$ cells.

The Hamiltonian we choose to describe the polymer chain has the form
$\hat H=\hat H_0+\hat H_U$ with
\begin{subequations}
\label{PTu1}
\begin{eqnarray}
\hat H_0&=&\sum_{\sigma,{\bf i}}\sum_{n,n'(n>n')}( t_{n,n'}
\hat c^{\dagger}_{{\bf i}+{\bf r}_n,\sigma} \hat c_{{\bf i}+{\bf r}_{n'},
\sigma} + H.c.) + \nonumber\\
&&+\sum_{\sigma,{\bf i}}\sum_{n=1}^m \epsilon_{n}
\hat n_{{\bf i}+{\bf r}_n,\sigma},\\
\label{PTu1a}
\hat H_U&=&\sum_{{\bf i}} \sum_{n=1}^m U_{n} \hat n_{{\bf i}+{\bf r}_n,
\uparrow} \hat n_{{\bf i}+{\bf r}_n,\downarrow}.
\label{PTu1b}
\end{eqnarray}
\end{subequations}
Here $\hat c^{\dagger}_{{\bf j},\sigma}$ creates an electron with spin $\sigma$
at site ${\bf j}$, $t_{n,n'}$ are hopping matrix elements connecting the sites
${\bf i}+{\bf r}_{n'}$ and ${\bf i}+{\bf r}_n$. Furthermore, $\epsilon_{n}$ and
$U_{n} > 0$ are on-site potentials and on-site Coulomb interactions,
respectively, defined at the sites ${\bf i}+{\bf r}_n$. The Fourier transform
of $\hat c_{{\bf i}+{\bf r}_n,\sigma}$ will be denoted by $\hat c_{n,{\bf k},
\sigma}$. We note that the Hamiltonian parameters
are arbitrary at this point, i.e., they are not chosen to provide
flat bands in the bare band structure. The case $\epsilon_n=0$ for all n
leaves the results qualitatively unchanged.

In the first step, we define $m-1$ block operators $\hat G^{\dagger}_{\alpha,
{\bf i},\sigma}=\sum_{\ell \in {\cal{B}}_{{\bf i},\alpha}} a_{\alpha,\ell}
\hat c^{\dagger}_{{\bf i}+{\bf r}_{\ell},\sigma}$ with $\alpha=1,\dots,m-1$,
i.e., linear superpositions of creation operators acting on blocks
${\cal{B}}_{{\bf i},\alpha}$ consisting of the m sites ${\bf i}+{\bf r}_{\ell}$
in the unit cell at ${\bf i}$; here $a_{\alpha,\ell}$ are numerical
coefficients. Specifically for the pentagon cell shown in Fig. 2 we employ 3
three-site blocks (the triangles made of sites (1,2,5), (2,3,5), (3,4,5)), and
2 two-site blocks (the site pairs (5,6), (4,7)) \cite{Generalization}.
The interaction term $\hat H_U$ is rewritten in terms of the operators
$\hat P_n = \sum_{\bf i} \hat P_{{\bf i}+{\bf r}_n}$, where
$\hat P_{\bf j}= \hat n_{{\bf j},\uparrow} \hat n_{{\bf j},\downarrow} -
(\hat n_{{\bf j},\uparrow} + \hat n_{{\bf j},\downarrow})+1$ is a positive
semidefinite operator with eigenvalue zero when there is at least one electron
on site ${\bf j}$. Altogether $\hat H-C_{g}$ takes the positive semidefinite
form $\hat H-C_{g}=\hat H_G + \hat H_P $, where
\begin{eqnarray}
\hat H_{G} = \sum_{\bf{i},\sigma} \sum_{\alpha=1}^{m-1} \hat G_{\alpha,{\bf i},
\sigma} \hat G^{\dagger}_{\alpha,{\bf i},\sigma},
\: \hat H_P = \sum_{n=1}^m U_{n} \hat P_n.
\label{PTu2}
\end{eqnarray}
Here $C_g=q_U N -N_c [\sum_{n=1}^m U_n + 2 \sum_{\alpha=1}^{m-1}z_{\alpha}]$,
$z_{\alpha}= \sum_{\ell}|a_{\alpha,\ell}|^2$, and $q_U$ are constants which
depend on the parameters entering in $\hat H$. The transformation of $\hat H$
into the semidefinite form shown in Eq. (\ref{PTu2}) requires that the
microscopic parameters in Eq. (\ref{PTu1}) fulfill certain conditions, i.e.,
equations connecting the coefficients $a_{\alpha,\ell}$ and $q_U$ to the
starting Hamiltonian parameters in Eq. (\ref{PTu1}). The solvability of the
matching conditions determines the parameter space domain ${\cal{D}}$ for which
the transformation from Eq. (\ref{PTu1}) to Eq. (\ref{PTu2}) can be performed.
It is not difficult to show that ${\cal{D}}$ is not strongly restricted by the
Hamiltonian parameters in Eq. (\ref{PTu1}). In particular, the values of the
interaction parameters $U_{n}$ can vary over a wide range. Details regarding the form of the matching conditions, the domain ${\cal{D}}$, and the block operators $\hat G^{\dagger}_{\alpha,
{\bf i},\sigma}$ in Eq. (2) are presented in the Appendix.

\begin{figure}[t!]
\epsfxsize=5.0cm
\centerline{\epsfbox{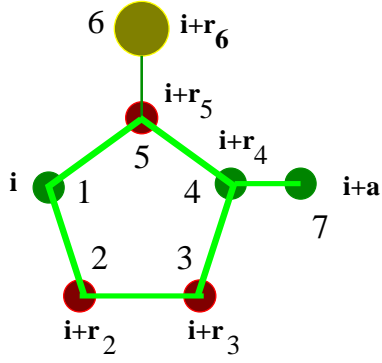}}
\caption{(Color online) The pentagon cell with $m_p=5$ and $m_e=1$. The numbers
indicate the index $n$ of the site, ${\bf a}$ is the primitive translation
vector, and ${\bf i}+{\bf r}_n$ specifies the site position inside the cell. }
\label{F1}
\end{figure}

Before we construct ground states of Eq. (\ref{PTu2}) above half filling
($\rho > 1/2$) we analyze its effective band structure. This is possible
because the anticommutation relations for the composite block operators
$\hat G^{\dagger}_{\alpha,{\bf i},\sigma}$, which depend on the interactions
$U_n$, allow us to rewrite the operator $\hat H_G$ as $\hat H_G =\hat H_{kin}+
K_G$, with a kinetic energy operator $\hat H_{kin}=-\sum_{\bf{i},\sigma}
\sum_{\alpha=1}^{m-1}\hat G^{\dagger}_{\alpha,{\bf i},\sigma}
\hat G_{\alpha,{\bf i},\sigma}$ and a constant $K_G=2N_c\sum_{\alpha=1}^{m-1}
z_{\alpha}$. The kinetic part, $\hat H_{kin}$, is quadratic in the original
fermionic operators $\hat c_{{\bf i}+{\bf r}_n,\sigma}$ and can hence be
diagonalized, leading to an effective, interaction dependent band structure.
In fact, the dispersion relations thereby obtained are identical to the energy
bands of $\hat H_0$ in Eq. (\ref{PTu1a}), but the on-site potentials
$\epsilon_{n}$ are now replaced by the renormalized energies
\begin{eqnarray}
\epsilon^R_{n}=\epsilon_{n} + U_{n} - q_{U}
\label{PTu3}.
\end{eqnarray}
It is this renormalization which can lead to an effective upper flat band. In
the ground state $|\Psi_g\rangle$ of $\hat H$ with $({\hat H}_G+{\hat H}_P)|
\Psi_g\rangle =0$ (see below), this flatness is unaffected by the presence of
$\hat H_P$ in Eq. (\ref{PTu2}) since $\hat H_P |\Psi_g\rangle=0$.

\begin{figure}[t!]
\epsfxsize=8cm
\centerline{\epsfbox{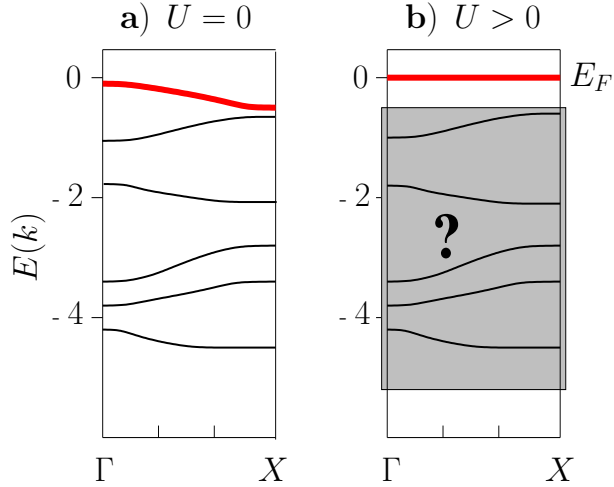}}
\caption{(Color online) Dispersion of the pentagon polymer chain.  a) Bare band structure (for details see the Appendix)
for hopping parameters $t_c\equiv t_{4,7}=0.5$, $t_h\equiv t_{3,2}=-1.1$,  $t_f\equiv t_{5,6}=1.2$,
and on-site potentials  $\epsilon_{1}=\epsilon_{4}=-2.5$, $\epsilon_{2}=\epsilon_{3}=-2.0$, $\epsilon_{5}=-2.1$, $\epsilon_{6}=-2.1$, using the site notation from Fig. 2. b) In the interacting case with the local interactions $U_{1}=U_{4}=0.18$, $U_{2}=U_{3}=0.35$, $U_{5}=U_{6}=0.03$ and the same $\hat H_0$ parameters as in a). The upper band, indicated by a thick line, is determined by Eq. (\ref{PTu3}).  Since $E(k)$ is even in $k={\bf k}\cdot{\bf a}$, only $k\in [0,\pi]$ is shown. The question mark in the shaded area indicates that the exact behavior in this region is not known. All energies are in units of $t\equiv t_{2,1}=t_{1,5}=t_{5,4}=t_{4,3}$, with $t>0$.}
\label{F2}
\end{figure}

Thus we find the very remarkable result that a dispersive band structure of
noninteracting electrons can be tuned by an interaction to yield an effective
upper flat band of the interacting, many-electron system; an example is given
Fig. \ref{F2} for a selected parameter set. The effectively flat band is half
filled for the total number of particles $N=2 N_{\Lambda}-N_c\equiv N^*$, and
is more than half filled for $N>N^*$. Such an interaction induced upper flat
band is possible only if the local interactions $U_{n}$ differ on at least one
site in the unit cell. We note that properties of the exact ground state can
only be deduced for the states in the upper band, the physics of the lower
bands is not accessible by the here applied method (as indicated by the
question mark in Fig. 3b).

\emph{The ground state for $N=N^*$}: In this case the ground state of Eq.
(\ref{PTu2}) has the form
\begin{eqnarray}
|\Psi_g(N^*)\rangle = [\prod_{\sigma} \hat G^{\dagger}_{\sigma} ]
\hat F^{\dagger} |0\rangle,
\label{PTu4}
\end{eqnarray}
where $|0\rangle$ is the vacuum state, $\hat G^{\dagger}_{\sigma}=\prod_{\bf i}
\prod_{\alpha=1}^{m-1}\hat G^{\dagger}_{m,{\bf i},\sigma}$, and the operator $\hat F^{\dagger}=\prod_{\bf i}\hat c^{\dagger}_{{\bf i}+
{\bf r}_{n_{\bf i}},\sigma}$ introduces one electron with spin $\sigma$ in each
unit cell. Since $\hat G^{\dagger}_{\sigma}$ creates $(m-1)N_c$ electrons with spin $\sigma$, the state $|\Psi_g(N^*)\rangle$ contains $N_{\sigma}=m N_c$ electrons with spin $\sigma$. Therefore there is one $\sigma$ electron on each site. Consequently all $\sigma$ electrons are localized, and only the $-\sigma$
electrons are mobile. Therefore Eq. (\ref{PTu4}) may be identically rewritten as
\begin{eqnarray}
|\Psi_g(N^*)\rangle=\prod_{\bf i} [(\prod_{n=1}^m\hat c^{\dagger}_{{\bf i}+
{\bf r}_n,\sigma})(\prod_{\alpha=1}^{m-1} \hat
G^{\dagger}_{\alpha,{\bf i}, -\sigma}) ] |0\rangle
\label{PTu5}.
\end{eqnarray}
 This state describes a half-metal, i.e., a non-saturated ferromagnet with
total spin $S=N_c/2$. Eq. (4) is indeed the ground state since
$\hat H_G |\Psi_g(N^*)\rangle=0$ and $\hat H_P|\Psi_g(N^*)\rangle=0$, where the
former relation is due $(\hat G^{\dagger}_{\alpha,{\bf i},\sigma})2=0$, while
the latter is a consequence of $N_{\sigma}=N_{\Lambda}$. Since every site is
occupied by an electron with spin $\sigma$ that part of the wave function which
describes the $-\sigma$ electrons is equivalent to a Slater
determinant.

The spatial extension of the electrons with spin $-\sigma$ is obtained from the long-distance ($r \to \infty$) behavior of the ground-state
expectation value of the hopping term $\Gamma_{\bf i}({\bf r})=\langle \Psi_g(N^*)|
(\hat c^{\dagger}_{{\bf i}+{\bf r}_n,-\sigma}
\hat c_{{\bf i}+{\bf r}_n +{\bf r},-\sigma}+ H.c.)|\Psi_g(N^*)\rangle $ for
arbitrary $n=1,\dots ,m$. Explicit calculations yield an exponential decay of $\Gamma_{\bf i}({\bf r})$ in the thermodynamic limit ($N_c\to\infty$). Hence the ferromagnetic state Eq. (5) is localized. Apart from the trivial $(2S+1)$ degeneracy related to the orientation of the total spin, where $S=S_z^{Max}=N_c/2$, the ground state \cite{AB} is unique; this was proved by us using the technique presented in detail in Ref. \cite{B23}.

\emph{The ground state for $N > N^*$}:
Here we restrict ourselves to the $S_z=S_z^{Max}$ sector. If we add $\bar N$
electrons to the system these electrons can occupy only $-\sigma$ spin states.
The ground state then has the form
\begin{eqnarray}
|\Psi_g(N^*+\bar N)\rangle = \hat Q^{\dagger}_{\bar N}
|\Psi_g(N^*)\rangle,
\label{PTu6}
\end{eqnarray}
where the operator $\hat Q^{\dagger}_{\bar N}= [\prod_{\gamma=1}^{\bar N}
\hat c^{\dagger}_{n_{\gamma},{\bf k}_{\gamma},-\sigma}]$ is a product of
$\bar N$ arbitrary, but different, $\hat c^{\dagger}_{n,{\bf k},-\sigma}$
operators. That Eq. (\ref{PTu6}) is indeed a ground state  follows from the
fact that $\hat G^{\dagger}_{\alpha,{\bf i},\sigma}$ anticommutes with the
fermionic creation operators $\hat c^{\dagger}_{n,{\bf k},-\sigma}$ and hence
also with the operators in $\hat Q^{\dagger}_{\bar N}$. Due to the free
electrons with spin $-\sigma$ introduced by $\hat Q^{\dagger}_{\bar N}$ the
ground state Eq. (\ref{PTu6}) contains also electrons in plane-wave like
states. Furthermore, 
since $E_g=C_g$,
one finds a vanishing charge excitation gap $\delta\mu =
E_{g}(N+1)-2E_{g}(N) +E_{g}(N-1)=0$ for $\bar N >1$.

In order to verify the extended character of the ground state Eq. (\ref{PTu6})
we calculate again the expectation value of the hopping term
$\Gamma_{\bf i}({\bf r})= \langle\Psi_g(N^*+\bar N)|
\hat c^{\dagger}_{{\bf i}+{\bf r}_2,-\sigma}
\hat c_{{\bf i}+{\bf r}_2+{\bf r}, -\sigma}+H.c. |\Psi_g(N^*+\bar N)\rangle$
specifically for $\bar N=1$ and $\hat Q_1=\hat c^{\dagger}_{2,{\bf k}^{(1)},
-\sigma}$, where ${\bf k}^{(1)}$ is the arbitrary momentum of the electron
added above $N=N^*$. In the thermodynamic limit of the pentagon chain the
result is
\begin{eqnarray}
\Gamma_{\bf i}({\bf r}) = \Gamma_0({\bf r}) (1- A({\bf k}^{(1)})/
B({\bf k}^{(1)})).
\label{gameq}
\end{eqnarray}
Here $\Gamma_0({\bf r})$ is the plane-wave result for a free electron in a
Bloch state, $A({\bf k})=A_1+A_2\cos{\bf k}\cdot{\bf a}$, and
$B({\bf k})=B_1+B_2\cos{\bf k}\cdot{\bf a} > 0$
holds for all ${\bf k}$ with constants $A_1$, $A_2$ and $B_1$, $B_2$ (for details 
see the Appendix).
This shows that the localization length is indeed infinite.

For pentagon chains without external links (i.e., when the sites ${\bf i}+
{\bf r}_6$ and ${\bf i}+ {\bf r}_7$ in Fig. 2 are missing) the same solutions are found, but the regions of parameter space where they exist are shifted.

We emphasize that in the case of Mielke-Tasaki ferromagnetism, i.e., flat-band
ferromagnetism in a half-filled lowest flat band \cite{MT}, both the flat band
and the connectivity conditions (the overlap of the local Wannier functions)
necessary for the emergence of this type of ferromagnetism, result from the
\emph{bare} band structure determined by $\hat H_0$. By contrast, we have shown here that the \emph{interaction} may be employed to tune a fully dispersive bare band structure to become partially flat. The results apply to the high-density region where Brocks \emph{et al.} conjectured the presence of strong correlations in acene and thiophene organic molecular crystals and the stabilization of magnetic phases \cite{B15}. Using the $\hat H_0$ parameters from Fig. 3, the matching conditions provide for the upper (bare) band a width of $W/t=0.15$, with $0.2 \leq U_n/W \leq 2.3$. For other parameters in $\hat H_0$  the parameter space domain ${\cal{D}}$ 
(see the Appendix) allows for even higher values of $U_{n}/W$.

Regarding the  experimental realization of the ferromagnetic and the
half-metallic states derived above we note that
the required electron doping of the pentagon chains can, in principle, be
achieved \cite{B15a} by two means: first by raising the Fermi level by
selecting appropriate side groups, and second by field-effect doping in a
double-layer transistor structure \cite{edlt1}.
Specifically for polythiophene with an estimated density of $10^{14}$ pentagon
rings/cm$^{2}$ \cite{edlt3}, electron densities of the order $10^{15}-10^{16}$
carriers/cm$^{2}$ are required. As already verified these electron densities
are experimentally achievable \cite{edlt5}.

In summary, by employing a rigorous analytic method we have constructed exact
ground states for a multiorbital pentagon Hubbard chain. The ferromagnetism and
the half-metallicity of the derived solutions originate from an unexpected
mechanism in multiorbital polygon chains with different site-dependent Coulomb
interaction strengths. For high electron densities with the top band at least
half-filled the interactions are capable of turning this dispersive band into
an effectively flat band in an extended parameter region. The obtained
solutions therfore point to a new route for the design of ferromagnetic
pentagon-chain polymers.

{\bf Acknowledgements.} We acknowledge valuable discussions with Wolfgang
Br\"utting on organic materials and with Michael Sekania on exact
diagonalization results. Support by the Alexander von Humboldt Foundation, the
TAMOP 4.2.1.-08/1-2008-003 research project of EC at the University of
Debrecen, and the Deutsche Forschungsgemeinschaft through SFB 484 in 2009 and
TRR 80 as of January 1, 2010,  is gratefully acknowledged.


\vspace*{1cm}

\hspace*{6cm}{\bf APPENDIX}

\section{1. Details regarding the transformed Hamiltonian, Eq. (2), for the pentagon chain}

a) The parameters of the transformed Hamiltonian, Eq.(2), obey the matching conditions
\begin{eqnarray}
&&\sum_{\alpha=1}^{m-1} z_{\alpha}=
4Q_1 + Q_1^2/|t_h| + 2 t^2/Q_1 + 2(|t_h|+|t_c|) + Q_3^2 + t_f^2/Q_3^2,
\nonumber\\ &&q_{U} = (1/2) [(U_{1}+\epsilon_{1}+|t_c|)+(U_{2}+\epsilon_{2}+
|t_h|) + \{ [(U_{2}+\epsilon_{2}+|t_h|)-(U_{1}+\epsilon_{1}+
|t_c|)]^2 +4t^2 \}^{1/2} ], \nonumber\\
&&Q_1=q_{U}-U_{2}-\epsilon_{2}-|t_h|, \quad Q_2=q_{U}-U_{1}-\epsilon_{1}-|t_c|,
\nonumber\\
&&Q_3 = |t_f|\sqrt{|t_h|}/[|t_h|(q_{U}-U_{5}-\epsilon_{5})-
(q_{U}-U_{2}-\epsilon_{2})^2 + t_h^2]^{1/2}, \hspace*{4.7cm}(A1)
\nonumber
\end{eqnarray}
where $Q_1,Q_2,Q_3 > 0$, and the hopping amplitudes $t, t_h, t_c, t_f$ are defined in the caption of Fig. 3.

\bigskip

b) The parameter space domain ${\cal{D}}$ for which the transformation from Eq.(1) to Eq.(2)
is valid is determined by the relations: \begin{eqnarray}
&&t_h<0, \quad Z=(q_{U}-Q_3^2) > \epsilon_{6},
\nonumber\\
&&W=q_{U}-[(q_{U}-U_{2}-\epsilon_{2})^2-t_h^2]/|t_h| > \epsilon_{5}, \nonumber\\
&&W-\epsilon_{5} > U_{5}>0, \quad U_{6}=Z-\epsilon_6. \hspace*{9.5cm}(A2)
\nonumber
\end{eqnarray}

c) The block operators entering in the transformed Hamiltonian, Eq.(2), are given by the
relations \begin{eqnarray}
&&\hat G_{1,{\bf i},\sigma}= q_S \hat b_{{\bf i},5,2} -
t / q_S  \hat c_{{\bf i}+{\bf r}_1,\sigma},
\nonumber\\
&&\hat G_{2,{\bf i},\sigma}= -t_S \hat b_{{\bf i},2,3} + Q_1
\hat c_{{\bf i}+{\bf r}_5,\sigma} / t_S,
\nonumber\\
&&\hat G_{3,{\bf i},\sigma}= q_S \hat b_{{\bf i},5,3} -
t / q_S  \hat c_{{\bf i}+{\bf r}_4,\sigma},
\nonumber\\
&&\hat G_{4,{\bf i},\sigma}= Q_3 \hat c_{{\bf i}+{\bf r}_6,\sigma} -
t_f/Q_3  \hat c_{{\bf i}+{\bf r}_5,\sigma},
\nonumber\\ &&\hat G_{5,{\bf i},\sigma}= |t_c|^{1/2} [\hat c_{{\bf i}+{\bf r}_4,\sigma} - t_c \hat c_{{\bf i}+{\bf a},\sigma} / |t_c| ], \hspace*{9cm}(A3)
\nonumber
\end{eqnarray}
where the notation $q_S=Q_1^{1/2}$, $t_S=|t_h|^{1/2}$,
$\hat b_{{\bf i},n,m}= \sum_{p=n,m} \hat c_{{\bf i}+{\bf r}_p,\sigma}$ was used.

\section{2. Dispersion of the bare bands of the pentagon chain presented in Fig. 3a}

The energy of the bare bands presented in Fig. 3a, $\epsilon=E_{\nu}(k)$, $\nu \leq m$, is provided by the equation
\begin{eqnarray}
&& 2 t_c t^2 [ (\epsilon_{6}-\epsilon) T_h - t_h T_f ] \cos k
+ T_f \{ t_h^2 t_c^2 - (\epsilon_{2}-\epsilon)^2 t_c^2  + [ (\epsilon_{1}-\epsilon)(\epsilon_{2}-\epsilon) - t^2]^2 - (\epsilon_{1}-\epsilon)^2 t_h^2 \} \nonumber\\
&& + 2(\epsilon_{6}-\epsilon) t^2 \{ t^2 (\epsilon_{2} + t_h - \epsilon)-(\epsilon_{1}-\epsilon) T_h \} = 0, \hspace*{7.2cm}(A4)
\nonumber
\end{eqnarray}
where $T_h=(\epsilon_{2}-\epsilon)^2-t_h^2$, $T_f=(\epsilon_{6}-\epsilon)
(\epsilon_{5}-\epsilon) -t_f^2$, and $k={\bf k} \cdot{\bf a}$.

\section{3. Details regarding the expectation value of the hopping term $\Gamma_{\bf i}({\bf r})$}

The parameters entering in the $\Gamma_{\bf i}({\bf r})$ expression from Eq.(7) are given by the relations
\begin{eqnarray}
&&A_1 =2|t_c|  \{  Q_3^2  [  2Q_1^2(2|t_n|+Q_1)+2t^2(Q_1+|t_n|) +
Q_1(t^2+2Q_1^2) + (Q_1^4+Q_1^2t^2)/|t_n| ] \nonumber\\
&&\hspace*{1cm}+
t_f^2/Q_3^2 [ t^2(|t_n|+Q_1) + Q_1^3 - (Q_1^4+Q_1^2t^2)/|t_n|] \},
\nonumber\\ &&A_2 = 2t_c t^2 [ (|t_n| t_f^2)/Q_3^2 - 2 Q_3^2(Q_1+|t_n|)],
\nonumber\\
&&B_1=2|t_c| \{  Q_3^2  [  (2|t_n|+Q_1^2/|t_n|)(Q_1^2+t^2) +
(t^2+2Q_1^2)(|t_n|+2Q_1) ] + (|t_n|t^2t_f^2)/Q_3^2 \},
\nonumber\\ &&B_2=2t^2t_c [ (|t_n|t_f^2)/Q_3^2 -Q_3^2(|t_n|+2Q_1) ], 
\hspace*{8.2cm}(A5)
\nonumber
\end{eqnarray}
and one has $B_1+B_2 \cos(k) > 0$ for all $k={\bf k}\cdot{\bf a}$.


\begin{references}
\bibitem{Generalization}
A similar strategy with qualitatively similar results can be applied for arbitrary polygons with $m_p$ sites in
the polygon and $m_e$ sites in side groups; in this case there are $m_p-2$
three-site and $m_e+1$ two-site blocks, leading to $m-1$ block operators.
\end{references}
\end{document}